\documentclass[preprint,12pt]{elsarticle}

\journal{Journal of Instrumentation}

\biboptions{sort&compress}
\begin{document}

\begin{frontmatter}

\title{A study on the radon removal performance of low background activated carbon}

\renewcommand{\thefootnote}{\fnsymbol{footnote}}
\author
{
Y.Y.~Chen$^{a}$,
Y.P.~zhang$^{b,c}$,
Y.~Liu$^{a}$,
J.C.~Liu$^{b,c}$,
C.~Guo$^{b,c}\footnote{Corresponding author. Tel:~+86-01088236256. E-mail address: guocong@ihep.ac.cn (C.~Guo).}$,
P.~Zhang$^{b,c}$,
S.K.~Qiu$^{a}$,
C.G.~Yang$^{b,c}$
Q.~Tang$^{a}\footnote{Corresponding author. Tel:~+86-1013974753537. E-mail address: tangquan528@sina.com (Q.~Tang).}$,
}
\address{
${^a}$ School of Nuclear Science and Technology, University of South China, Hengyang, China\\
${^b}$ Experimental physics division, Institute of High Energy Physics, Chinese Academy of Science, Beijing, China\\
${^c}$ School of Physics, University of Chinese Academy of Science, Beijing, China\\
}

\begin{abstract}
Radon and radon daughters pose significant backgrounds to rare-event searching experiments. Activated carbon, which has very strong adsorption capacity for radon, can be used for radon removal and radon enrichment. The internal $^{226}$Ra concentration ultimately limits its radon enrichment ability. In order to measure the intrinsic background and study the radon adsorption capability of Saratech activated carbon at various temperatures, a radon-emanation measurement system with a high-sensitivity radon detector and an adsorption-performance research-system have been developed. In this paper, a 0.71~mBq/m$^3$ high-sensitivity radon detector and measurement details of the radon-adsorption capability of Saratech activated carbon at low temperature will be presented.
\end{abstract}


\end{frontmatter}

%\linenumbers

\section{Introduction}
Neutrino experiments and dark matter searching experiments are two important directions of particle physics, which have very strict requirements on detector background. $^{222}$Rn and its progenies are one of the most crucial intrinsic background sources, especially for those with liquid noble elements (such as LZ~\cite{LZ}, PandaX~\cite{PandaX} and DarkSide~\cite{DarkSide}), liquid scintillator(such as JUNO~\cite{JUNO} and Borexino~\cite{Borexino}) and ultra-pure water (such as Super-K~\cite{SuperK1}) as target materials. For example, to the dark matter searching experiments based on xenon, backgrounds induced by $^{214}$Pb, $^{214}$Bi, $^{210}$Pb and $^{210}$Bi, four $\beta$-decaying isotopes in $^{222}$Rn decay chain, dominate in the low energy region~\cite{Xenon1}. Apart from the air leakage of the detector, $^{222}$Rn, which is the daughter isotope of $^{226}$Ra in $^{238}$U decay chain, can continuously emanate from detector materials. As a noble gas with a half life of 3.8 days, $^{222}$Rn can enter the target by means of diffusion or recoil from prior $\alpha$-decays. Once in the liquid, $^{222}$Rn distributes homogeneously, reaching also the innermost part of the detector. In order to remove radon from the detector target, many experiments have setup online radon removal equipments, such as Super-K~\cite{SuperK2} and Xenon100~\cite{Xenon2}.

Activated carbon is an effective adsorbent for various impurities by physical adsorption and its radon adsorption capability has been validated by many experiments. The Super-K collaboration used it to produce radon free air~\cite{SuperK1}, the DarkSide collaboration used it to remove radon from argon~\cite{DarkSide} and the XMASS collaboration used it to remove radon from xenon~\cite{XMASS}. For activated carbon used in low background experiments, it is necessary to reduce the internal radioactive impurities because radon emanated from the carbon itself ultimately limits its radon adsorption capability. Carboact activated carbon~\cite{Carboact} has the lowest background ($\sim$ 0.3~mBq/kg~\cite{charcoal,charcoal2}) but the price is extremely expensive (15000~USD/kg)~\cite{charcoal}. Saratech activated carbon~\cite{Saratech} is a good alternative for many physicists because its background is comparable to Carboact ($\sim$1~mBq/kg~\cite{charcoal}) while the price is much cheaper (35~USD/kg)~\cite{charcoal}. In order to study the application prospect of Saratech activated carbon in low background experiments, a radon-emanation system has been developed to measure the background, and an adsorption-performance research-system has been developed to study its radon adsorption capability at various temperatures.

This paper is organized as follows. In section 2, an overview of the two systems will be described. In section 3, the detail of the radon detector, including  optimization, calibration and sensitivity estimation, will be described. In section 4, the intrinsic background of Saratech activated carbon and its radon adsorption capability at various temperatures will be described. In section 5, the conclusions and application prospect will be presented.

\section{Experimental setup}
\subsection{Radon-emanation measurement system}
\label{sec.2}

\begin{figure}[htb]
\centering
\includegraphics[width=10cm]{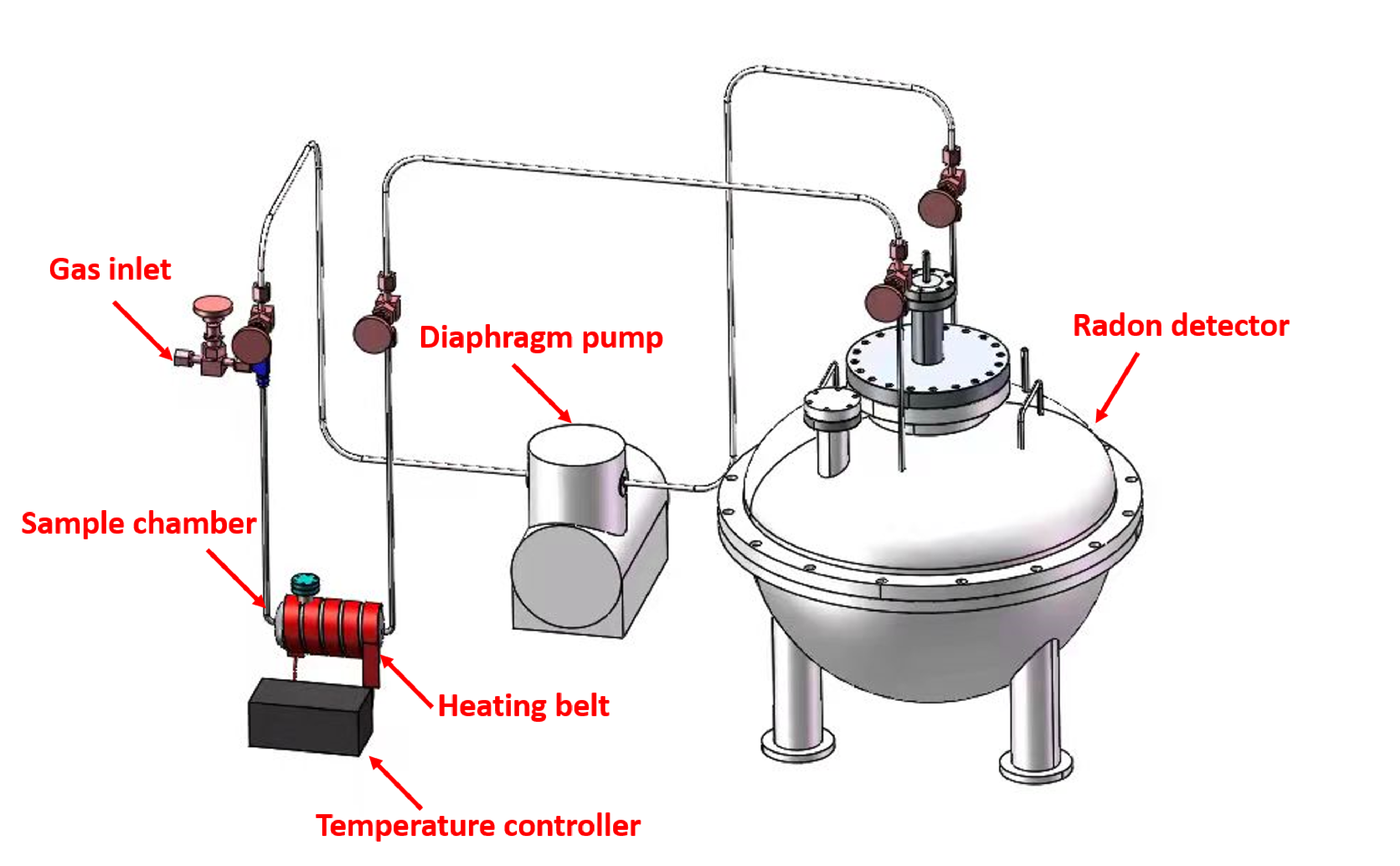}
\caption{Scheme of the radon-emanation measurement system, which consists of a high-sensitivity radon detector, a sample chamber, a diaphragm pump and a temperature control system.}
\label{fig.wholesystem}
\end{figure}

A radon-emanation measurement system has been developed to measure the intrinsic $^{226}$Ra background of the activated carbon. Fig.~\ref{fig.wholesystem} shows the scheme of the system, which consists of a high-sensitivity radon detector, a sample chamber, a diaphragm pump and a temperature control system.

\begin{figure}[htb]
\centering
\includegraphics[width=12cm]{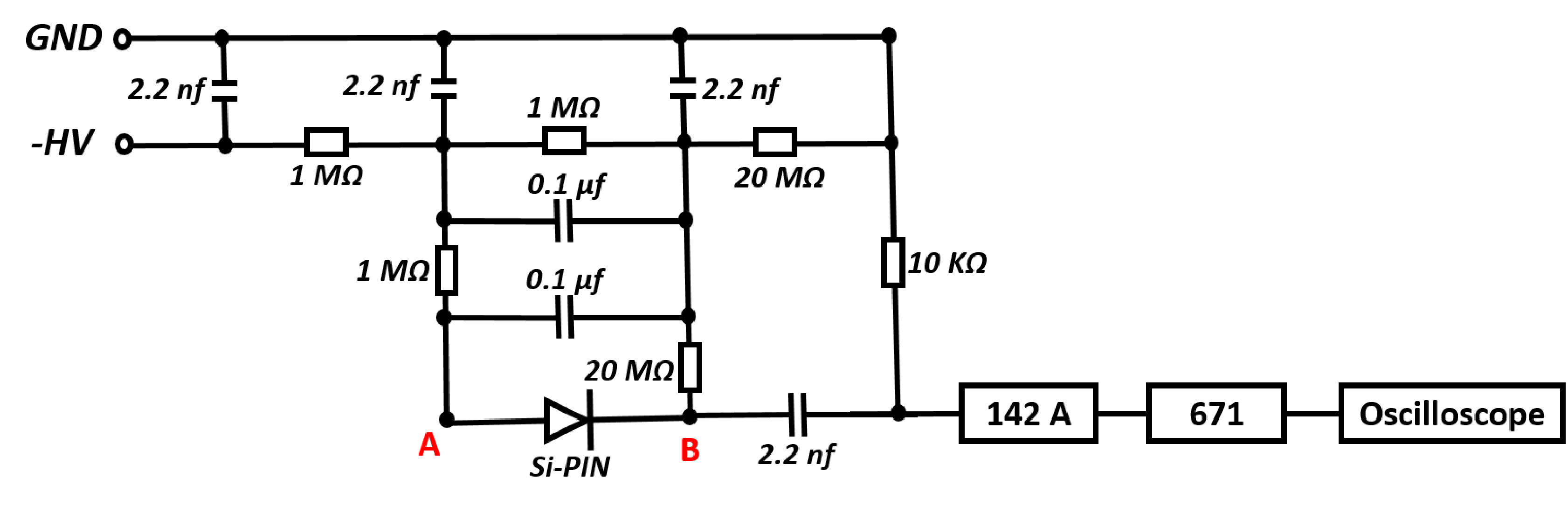}
\caption{The circuit diagram of the readout system. 142A is a pre-amplifier and 671 is a main amplifier made by ORTEC company. If -HV is set to -1000 V, then V$_A$ $\approx$ -955~V, V$_B$$\approx$-910~V, V$_{bias}$$\approx$45~V. }
\label{fig.circuitdiagram}
\end{figure}

(A) The high-sensitivity radon detector is used to determine the radon concentration in the gas. Compared with our previous work~\cite{Rn_2018,Rn_2020}, the geometry of the detector has been optimized to hemispherical for better collection efficiency and the loading mode of the collecting electric field has been changed from positive high voltage to negative high voltage so as to reduce the materials inside and obtain a lower detector background. The circuit diagram is shown in Fig.~\ref{fig.circuitdiagram}, the potentials at both ends and the bias of the Si-PIN is illustrated in the caption. The volume of the detector is 41.5~L and only a Si-PIN photodiode (S3204-09, Hamamatsu) is placed inside. The scheme of the readout system is shown in Fig.~\ref{fig.circuitdiagram}. Signals from Si-PIN are amplified by a pre-amplifier (142A, ORTEC) and a main amplifier (671, ORTEC) before being recorded by an oscilloscope (610Zi, LeCroy).

\begin{figure}[htb]
\centering
\includegraphics[width=4cm]{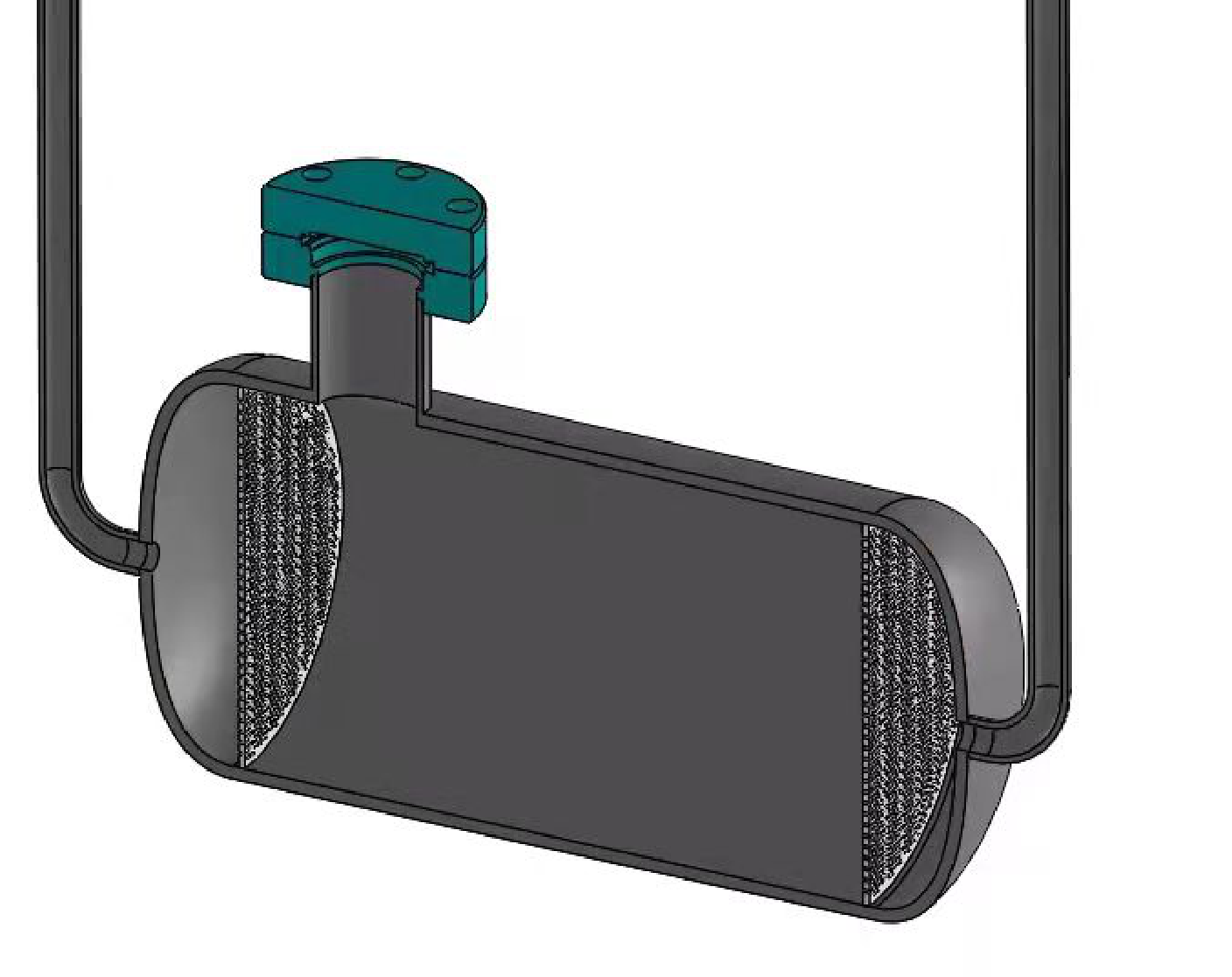}
\caption{Scheme of the sample chamber. Stainless steel sintering meshes are welded at both ends of the chamber. }
\label{fig.samplechamber}
\end{figure}

(B) The sample chamber, with a volume of 0.2~L, is used to place the measured sample. This chamber is specially designed for measuring the $^{226}$Ra background of activated carbon. The detail of the chamber is show in Fig.~\ref{fig.samplechamber}. Two pieces of stainless steel sintering meshes are welded at both ends to prevent activated carbon from being blown out by the gas flow. The aperture size of the meshes is $\sim$0.2~mm.

(C) The diaphragm pump (N022AT.18, KNF) is used for gas circulation because it is difficult to keep the radon concentration in the sample chamber consistent with that in the detector by diffusion. The connectors of the pump have been changed from G 1/8 to VCR 1/4 to get a lower air leakage rate.

(D) The temperature control system is used for radon desorption and consists of a temperature controller, a temperature sensor (Pt100) and a heating belt wound around the sample chamber. The Pt100, which is placed in the gap between the heating belt and the sample chamber, serves as the input to the temperature controller and the stability of the system is within $\pm$3~$^\circ$C.

In order to get a lower background, the radon detector, the sample chamber and all the pipelines in the system are electro-polished to a roughness of 0.1~$\mu$m. Knife-edge flanges with metal gaskets as well as VCR connectors with metal gaskets are used. The leakage rate of the system is better than 1$\times$10$^{-8}$~Pa$\cdot$m$^3$/s, which is measured by a helium leak detector (ZQJ-3000, KYKY Technology Co. Ltd).

\subsection{Adsorption-performance research-system }

\begin{figure}[htb]
\centering
\includegraphics[width=10cm]{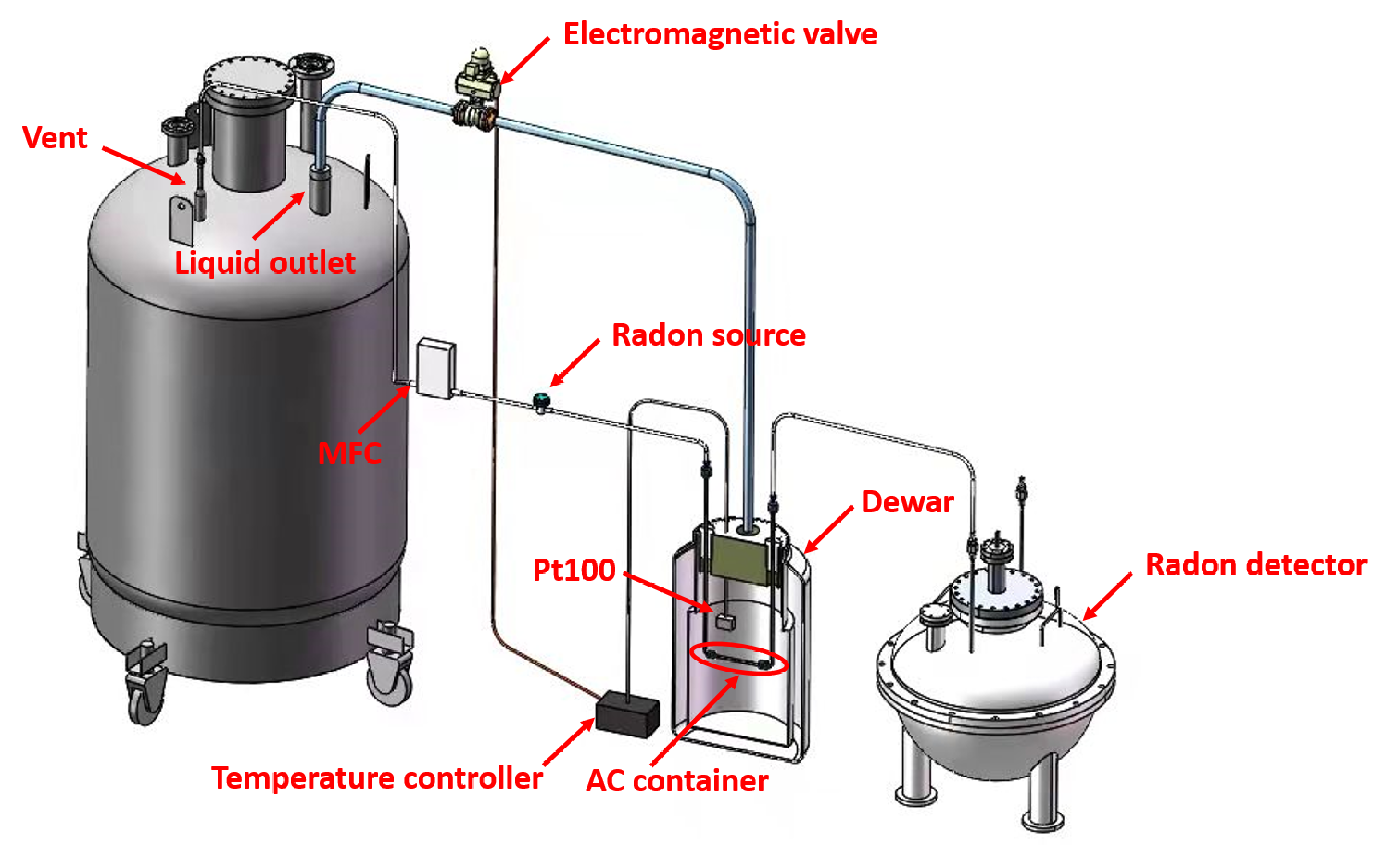}
\caption{Scheme of the adsorption-performance research-system, which consists of a liquid nitrogen tank, a mass flow control, a solid radon source, a dewar, an activated carbon container, a temperature control system and a radon detector.}
\label{fig.charcoalsystem}
\end{figure}

In order to study the radon adsorption-performance of activated carbon, an adsorption-performance research-system, which is shown in Fig.~\ref{fig.charcoalsystem}, has been developed. The system consists of a liquid nitrogen tank, a mass flow controller (MFC), a solid radon source, an activated carbon container, a temperature control system and a radon detector.

(A) The liquid nitrogen tank is filled with liquid nitrogen. The vent of the tank is used to supply evaporated nitrogen and the relative pressure inside is $\sim$1 bar. The liquid outlet of the tank is used to supply liquid nitrogen to the temperature control system.

(B) The MFC (1179A, MKS) is used to control the flow rate of the gas, so as to keep the radon concentration in the gas stable.

(C) The radon source used in this work is similar to the one that was used in our previous work~\cite{Rn_2020}, while the container has been changed to stainless steel with VCR connectors. The radon concentration in the gas is inversely proportional to the flow rate.

(D) The activated carbon container is a 15~cm long 1/4 VCR pipeline. Two filter gaskets are placed at both ends. The aperture size of the gaskets is 0.2~mm. The container is placed in a dewar and the temperature inside the dewar is controlled by the temperature control system.

(E) The temperature control system consists of a dewar, a temperature controller, a temperature sensor (Pt100) and an electromagnetic valve. The Pt100 temperature sensor which is placed next to the activated carbon chamber serves as the input to the temperature controller. The temperature controller controls the switch of the electromagnetic valve, so as to realize the intermittent injection of liquid nitrogen into the dewar. The liquid nitrogen vaporize in the dewar and the cooled nitrogen gas can cool down the activated carbon. This system can realize a temperature change inside the dewar from room temperature to -196~$^\circ$C and the temperature stability is within $\pm$5~$^\circ$C.

(F) The radon detector used in this measurement system is the same as the one described in Sec.~\ref{sec.2}.

\section{The radon detector}
\subsection{Detection principle}
\label{sec.DP}

\begin{figure}[htb]
\centering
\includegraphics[width=10cm]{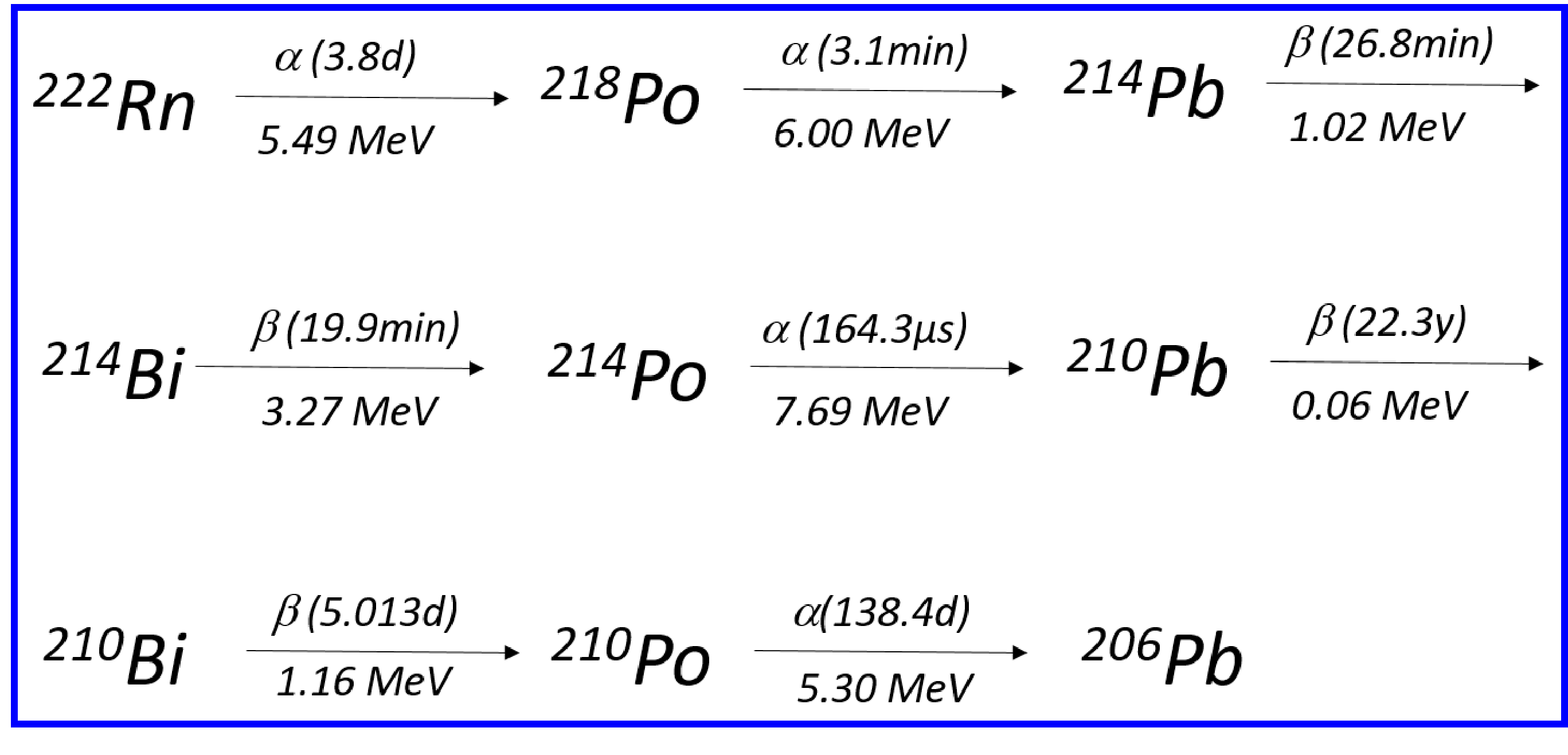}
\caption{The relevant branches of $^{222}$Rn decay chain. The time in the bracket is the half-life of the nuclide. The energy for $\beta$ decay refers to the maximum one.}
\label{fig.decaychain}
\end{figure}

\begin{figure}[htb]
\centering
\includegraphics[width=6.6cm,height=4cm]{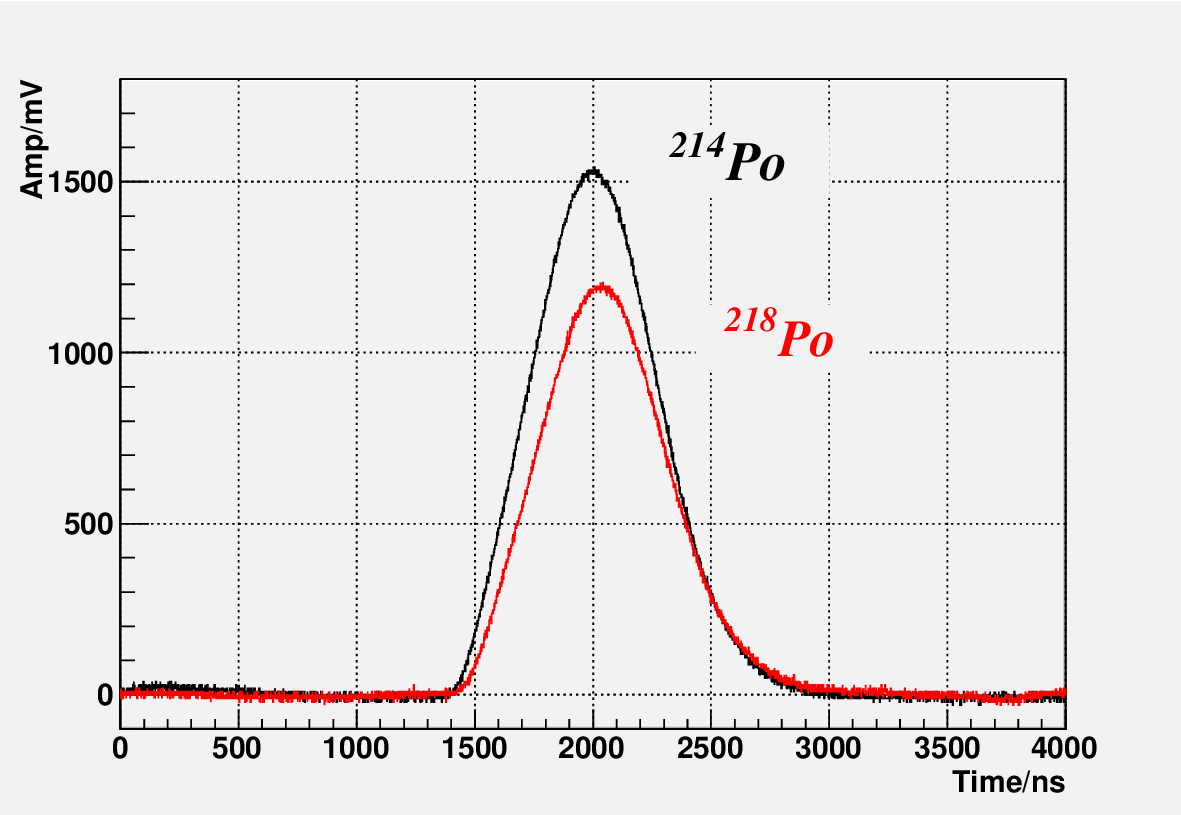}
\includegraphics[width=6.6cm,height=4cm]{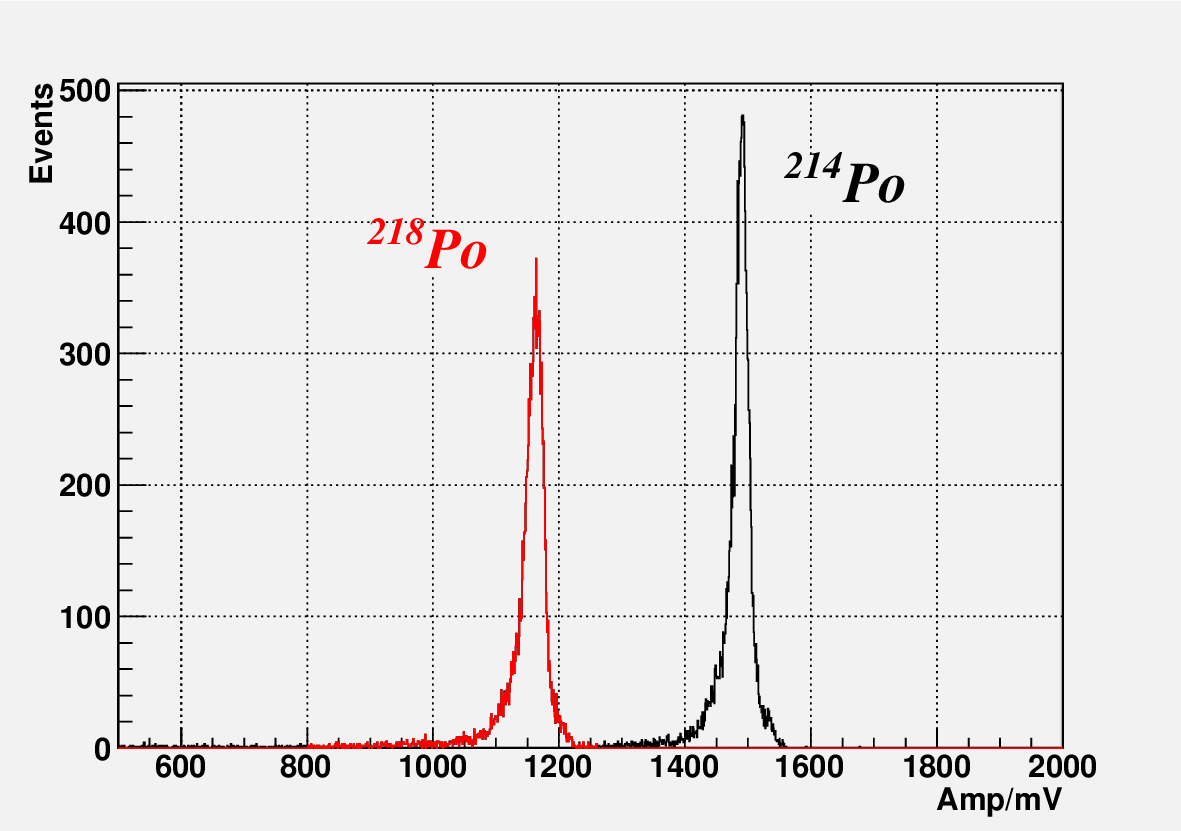}
\caption{Left: Examples of signal pulses, the black curve is an example pulse of $^{214}$Po and the red curve is an example pulse of $^{218}$Po. Right: Spectrum of $^{222}$Rn source, the red peak is caused by $^{218}$Po and the black peak is caused by $^{214}$Po. }
\label{fig.sample}
\end{figure}

The radon detector consists of a hemispherical chamber, a Si-PIN photodiode and supporting electronics. The basic principle of the detector is electrostatic collection. Fig.~\ref{fig.decaychain} shows the relevant branches of $^{222}$Rn decay chain. After $\alpha$ decay of $^{222}$Rn, its daughter nucleus will obtain a large recoil energy and then lose part of the electrons in the outer layer. Therefore, more than 90\% of $^{222}$Rn daughters are positively charged~\cite{positive}, so they can be collected to the surface of the Si-PIN photodiode by the electric field. By detecting $\alpha$s from $^{218}$Po and $^{214}$Po decay, the $^{222}$Rn concentration inside the chamber can be determined. The window of the Si-PIN photodiode is unsealed, so that the $\alpha$s, which have very weak penetration ability can be detected with very high efficiency while $\beta$s or $\gamma$s can not be detected because the thickness of the Si-PIN sensitive area is very thin and the energy deposited by $\beta$s or $\gamma$s on it is very small~\cite{SiPIN} . By using the circuit shown in Fig.~\ref{fig.circuitdiagram}, we can realize that the Si-PIN has a positive bias while is itself at a negative potential. The shell of the stainless steel chamber is grounded, so an electrical field is generated from the chamber shell to the Si-PIN. Before being recorded by an oscilloscope, signals are amplified by a pre-amplifier and a main amplifier. The left of Fig.~\ref{fig.sample} shows example pulses of $^{218}$Po and $^{214}$Po and the right of Fig.~\ref{fig.sample} shows the energy spectrum of a $^{222}$Rn source. The event rate of $^{214}$Po is used to calculate the radon concentration because there are no other $\alpha$ sources in its signal region~\cite{SuperK3} and $^{214}$Po has a higher collection efficiency than $^{218}$Po, which also can be seen from the right plot of Fig.~\ref{fig.sample}.

\subsection{Detector calibration}
In order to get the corresponding relationship between $^{214}$Po event rate and $^{222}$Rn concentration in the detector, a gas flow $^{222}$Rn source, which is made from BaRa(CO$_3$)$_2$ powder by the radon laboratory of South China university, has been used to calibrate the detector. During the calibration, the gas flow rate is set to 1~L/min, which is controlled by the MFC, and the radon concentration in the outgas is 33.0$\pm$3.3~Bq/m$^3$, which is measured by a RAD7 radon detector of Durridge Company Inc~\cite{RAD7}. A calibration factor (C$_F$) is defined as Eq.~\ref{Eq.CF}:

\begin{equation}
C_F[(counts/h)/(Bq/m^3)]= \frac{measured~  ^{214}Po~signal~rate}{^{222}Rn~concentration}
\label{Eq.CF}
\end{equation}

where the numerator is the event rate of $^{214}$Po in the unit of counts per hour (cph), and the denominator is the $^{222}$Rn concentration in the unit of Bq/m$^3$. For a radon detector, the internal radon activity can be calculated according to the radon concentration and the detector volume, and then the maximum event rate can be obtained. Therefore, the value of C$_F$ also represents the detection efficiency of the detector.

\begin{figure}[htb]
\centering
\includegraphics[width=7cm]{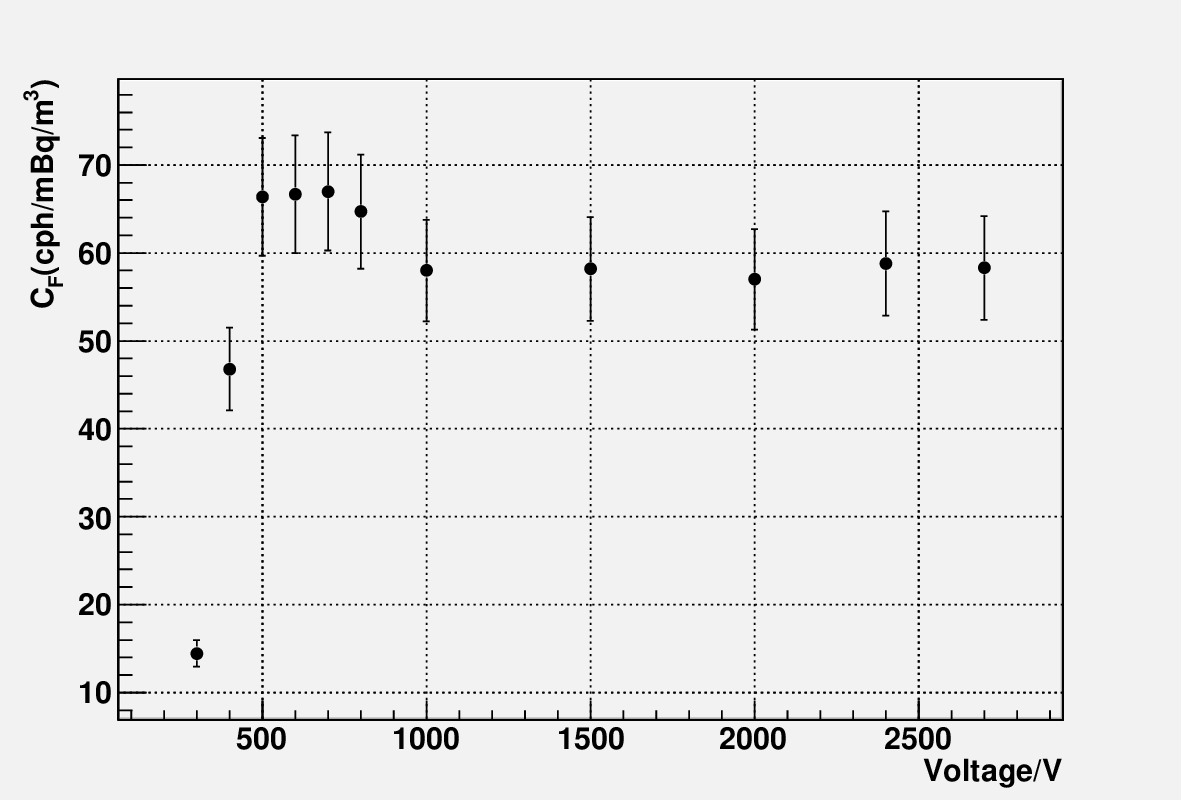}
\caption{The variation of C$_F$ with applied voltage. C$_F$ is calculated according to Eq.~\ref{Eq.CF} and the uncertainties include both systematic uncertainties and statistical uncertainties. The systematic uncertainties mainly come from the fluctuation of the radon concentration in the gas. }
\label{fig.CF}
\end{figure}

According to the detection principle described in sec.~\ref{sec.DP}, the detection efficiency is related to the humidity and the electric field inside the detector. Because the gas source in the experiment is evaporated nitrogen and the absolute humidity of evaporated nitrogen is $\sim$0.01~$\mu$g/cm$^3$, only the influence of the electric field on the detection efficiency is considered. The variation of C$_F$ with the applied voltage is shown in Fig.~\ref{fig.CF}.

As can be seen from Fig.~\ref{fig.CF}, C$_F$ reaches the maximum value at $\sim$700V, which is 67.0$\pm$6.7~cph/(Bq/m$^3$). This corresponds to a collection efficiency of $\sim$90\%. Compared with our former work~\cite{Rn_2018}, this improvement mainly comes form the geometry optimization of the detector. The operation voltage is set to 700~V in the following experiment.

\subsection{Sensitivity estimation}

\begin{figure}[htb]
\centering
\includegraphics[width=7cm]{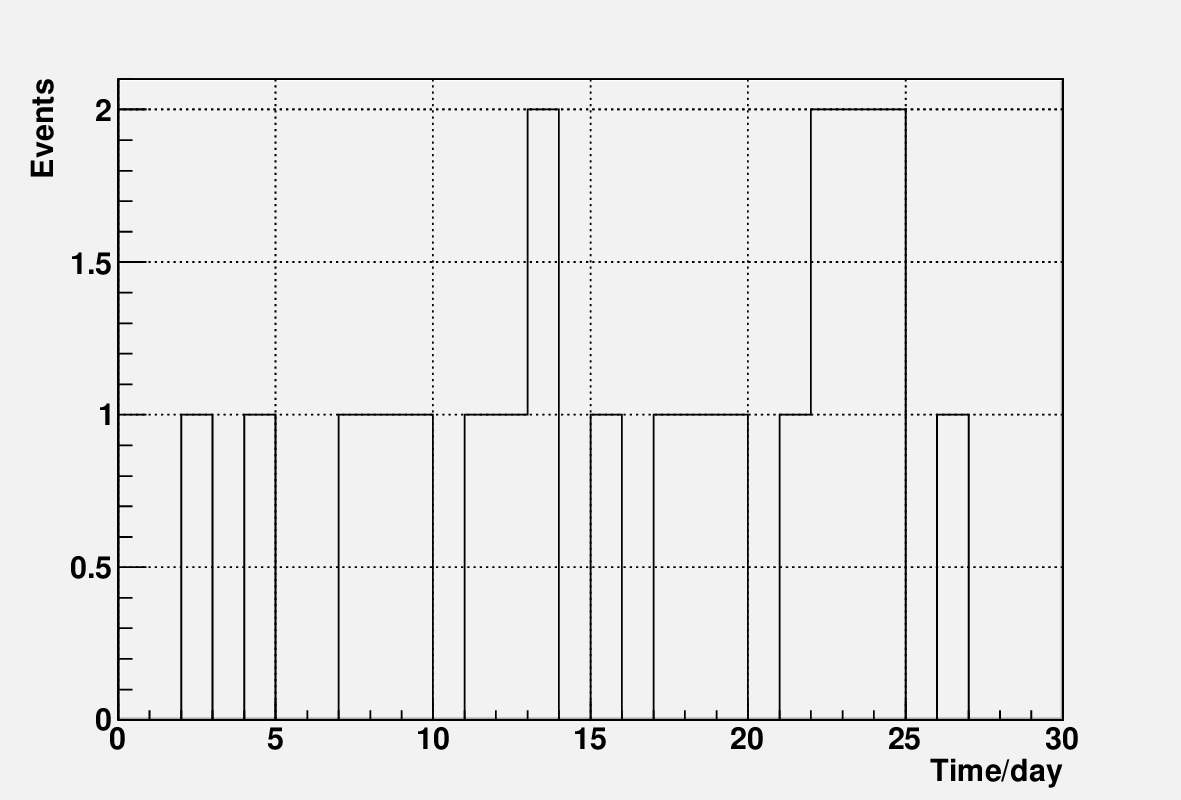}
\caption{The background event rate of the detector.}
\label{fig.background}
\end{figure}

The intrinsic background of the detector is measured to estimate the sensitivity. Fig.~\ref{fig.background} shows the background event rate for 30 days. Before measuring the background, the detector has been evacuated with a vacuum pump (ACP40, Pfeiffer) and flushed with evaporated nitrogen for more than 10 volumes. The averaged background event rate is 0.70 $\pm$ 0.15 (statistical uncertainty only) counts per day (cpd). At 90\% confidence level, the characteristic detection limit of the detector can be estimated according to Eq.~\ref{Eq.sensitivity}~\cite{SuperK3}:

\begin{equation}
L_c = 1.64 \times \sigma_{BG}/C_F
\label{Eq.sensitivity}
\end{equation}

where L$_c$ is the detector sensitivity, $\sigma_{BG}$ is the uncertainty on the background event rate, C$_F$ is the calibration factor. From the background measurement, a sensitivity of 0.71~mBq/m$^3$ is derived for one day measurement.

\section{Low background activated carbon}

\begin{figure}[htb]
\centering
\includegraphics[width=7cm]{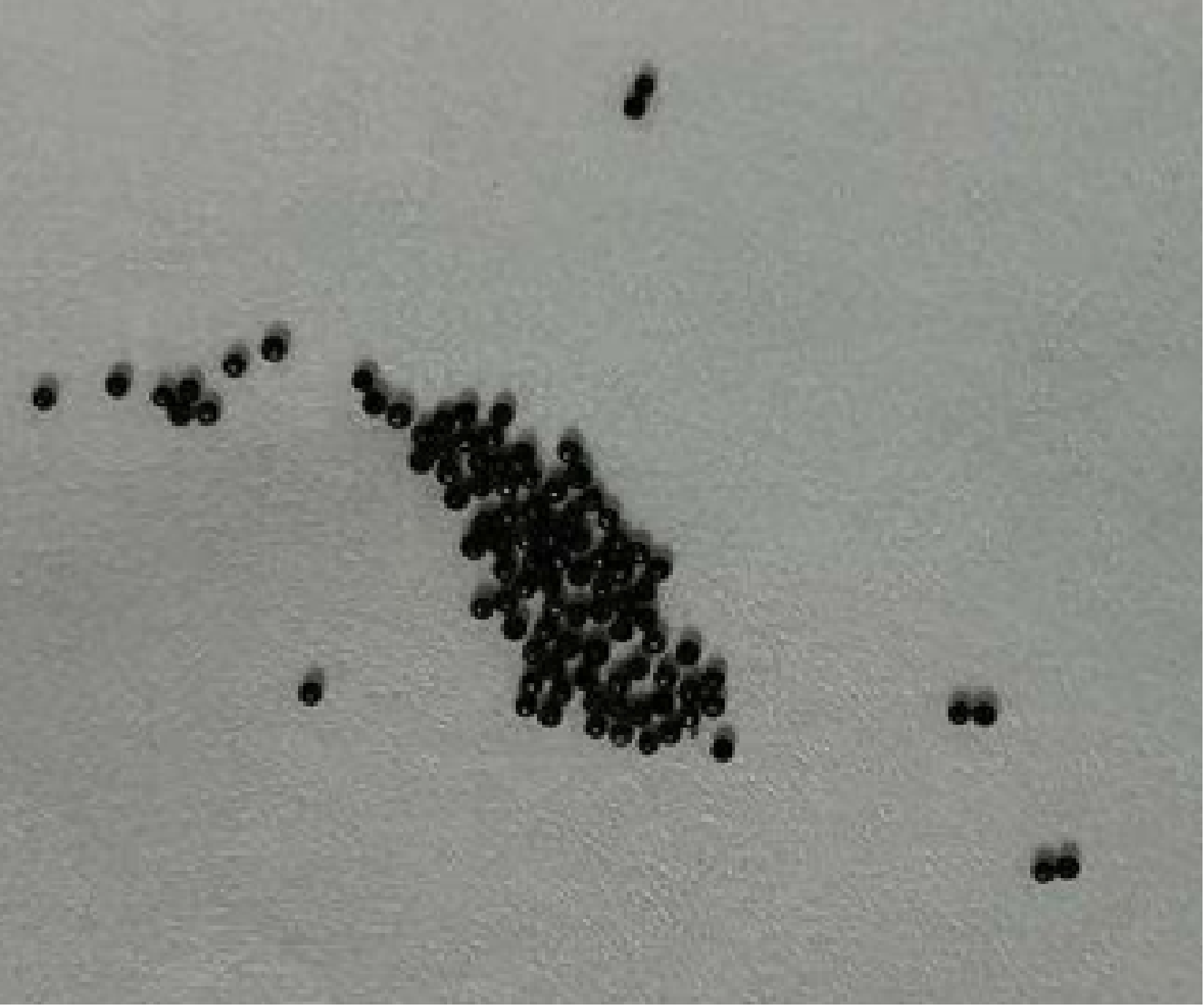}
\caption{Picture of Saratech activated carbon.}
\label{fig.saratech}
\end{figure}

Activated carbon has been widely used in low background experiments to remove the radon in the detector. The adsorption capability of activated carbon depends on its micro-porous structure, chemical composition, various functional groups and non-carbon atoms on the carbon surface. Apart from the radon adsorption capability, the intrinsic background is another factor that has to be taken into consideration for low background experiments. According to Ref.~\cite{charcoal}, the activated carbon made by Saratech company, with a surface of 1.34~$\times$~10$^7$~cm$^2$/g and a background of $\sim$1~mBq/kg, is a good choice. Fig.~\ref{fig.saratech} shows a picture of Saratech activated carbon.

\subsection{Background measurement}
The intrinsic background of activated carbon may vary from batch to batch and the carbon may also be polluted during packaging and transportation, thus a measurement is necessary before use. 137~g of activated carbon have been used for background measurement with the system shown in Fig.~\ref{fig.wholesystem}. No special treatment has been done before measuring and the measurement procedure is as follows:

(A) Put the activated carbon in the sample chamber and heat the activated carbon to 180~$^\circ$C.

(B) Purge the heated activated carbon with evaporated nitrogen for 2 hours to remove the adsorbed radon from the activated carbon. The flow rate of the nitrogen is set to 3~L/min.

(C) Seal the sample chamber for $^{222}$Rn accumulation and turn off the temperature control system. In this measurement, the sample chamber is sealed for 139~hours.

(D) Turn on the temperature control system and turn on the pump to transfer the radon gas from the sample chamber to radon detector after the temperature stabilized at 180~$^\circ$C.

(E) Isolate the radon detector after 2 hours gas circulation.

(F) Calculate the $^{226}$Ra concentration with

\begin{equation}
C_{Ra} = \frac{(n-n_0) \times (V_D+V_C)}{m \times C_F \times 24 \times (1-e^{-\lambda_{Rn} \times t})}
\label{Eq.CRa}
\end{equation}

where C$_{Ra}$ is the radium concentration of the activated carbon in the unit of Bq/kg, n and n$_0$ are the signal event rate and background event rate respectively in the unit of cpd, V$_D$ is the detector volume in the unit of m$^3$, V$_C$ is the volume of sample chamber in the unit of m$^3$, m is the mass of activated carbon in the unit of kg, C$_F$ is the calibration factor in the unit of cph/(Bq/m$^3$), $\lambda_{Rn}$ is the decay constant of $^{222}$Rn, t is the sealing time of the sample chamber. The values of the variables of Eq.~\ref{Eq.CRa} are listed in Tab.~\ref{tab:CRa} and the $^{226}$Ra concentration in the activated carbon is 1.36 $\pm$ 0.94~mBq/kg.

\begin{table}[htb]
\begin{center}
\caption{Values of the variables used in Eq.~\ref{Eq.CRa}.}
\begin{tabular}[c]{c|cccc} \hline
Var & n~(CPD) & n$_0$~(CPD) & V$_D$~(m$^3$) & V$_C$~(m$^3$) \\\hline
Val & 9$\pm$3 & 4.3$\pm$1.2 & 0.0415 & 0.0002 \\\hline
Var& m~(kg) & C$_F$ (CPH/(Bq/m$^3$)) & $\lambda_{Rn}$~(s$^{-1}$) & t~(s) \\\hline
Val& 0.137 & 67.0$\pm$6.7 & 2.1e$^{-6}$ & 141*60*60 \\\hline
\end{tabular}
\label{tab:CRa}
\end{center}
\end{table}

\subsection{Radon adsorption performance}
Activated carbon has strong adsorption capability for radon, especially at low temperature. The system shown in Fig.~\ref{fig.charcoalsystem} is used to quantitatively study the adsorption capability of activated carbon. According to Ref.~\cite{XMASS}, 1~g activated carbon can be used to remove radon from more than 1~m$^3$ nitrogen. In order to save time, 0.73~g activated carbon are used in this experiment. The operation procedure is as follows:

(A) Heat the activated carbon to 180~$^\circ$C and purge it with evaporated nitrogen for 2 hours.

(B) Turn on the temperature control system to cool down the activated carbon to the specified temperature.

(C) Open the vent of the liquid nitrogen tank and set the flow rate to 1~L/min, at which the radon concentration in the gas is 33 $\pm$ 3.3~Bq/m$^3$.

(D) Monitor the radon concentration inside the radon detector until equilibrium is reached.

During this test, the event rate of $^{218}$Po, whose half life is 3.1~min, is used to monitor the radon concentration inside the detector. $^{214}$Po itself takes $\sim$2 hours to reach equilibrium, which will introduce a large uncertainty on the measurement of the saturation time of activated carbon.

\begin{figure}[htb]
\centering
\includegraphics[width=7cm]{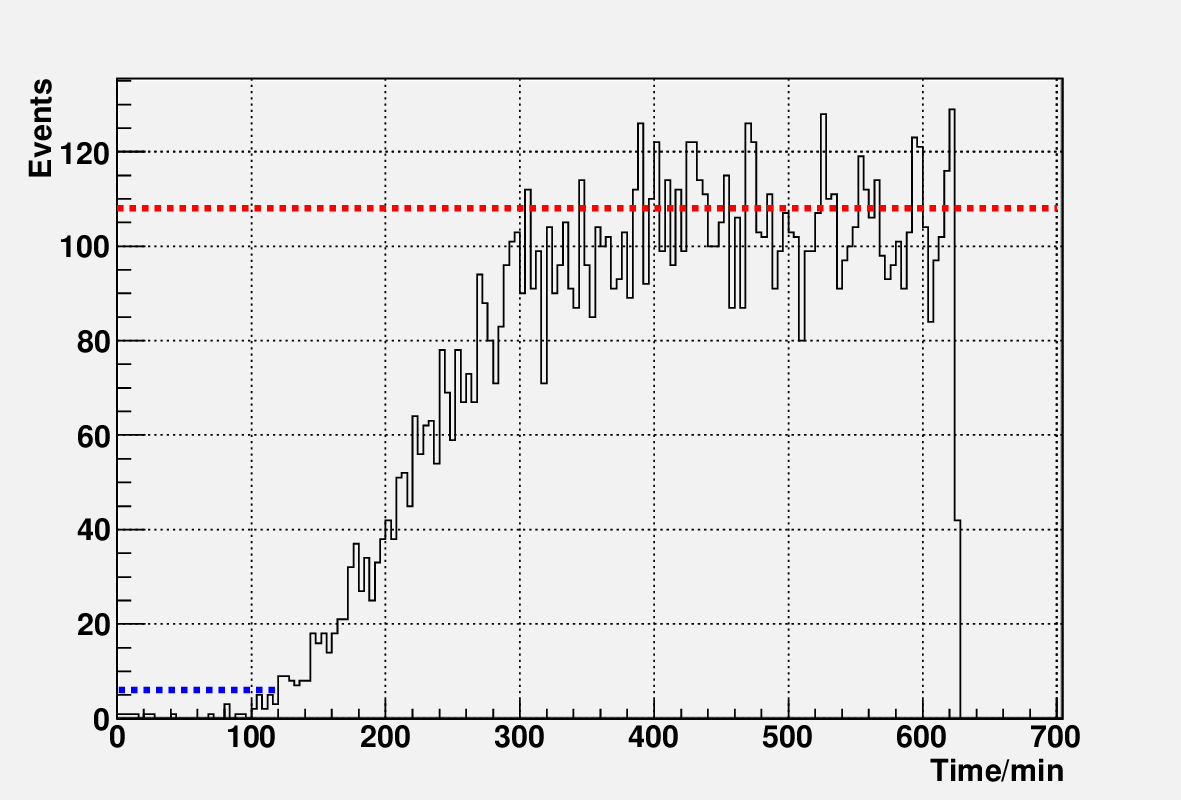}
\caption{Changes of $^{218}$Po event rate with time inside the detector at -60~$^\circ$C. The average event rate at equilibrium is indicated with red dashed line and 5\% event rate is indicated with the blue dashed line.}
\label{fig.charcoal60}
\end{figure}

Whether using activated carbon for radon removal or radon enrichment, the radon adsorption efficiency is a factor that has to be considered. In this study, an efficiency of 95\% is used as the baseline. Fig.~\ref{fig.charcoal60} shows the changes of $^{218}$Po event rate along with time inside the detector at -60~$^\circ$C. During the first $\sim$100 minutes, activated carbon has almost 100\% adsorption efficiency and all the radon is trapped. From $\sim$100 minutes to $\sim$320 minutes, the efficiency decreases due to the gradual saturation of activated carbon and at $\sim$320 minutes, the efficiency decreased to zero. In Fig.~\ref{fig.charcoal60}, the average event rate at equilibrium is indicated with the red dashed line and 5\% event rate is indicated with the blue dashed line. The intersection of the blue dashed line and the experimental data indicates that the radon adsorption efficiency is 95\% after $\sim$122~minutes. The radon adsorption capability is defined as:

\begin{equation}
Q = V_g \times t_s/m_{AC}
\label{Eq.Q}
\end{equation}

where Q is the radon adsorption capability in the unit of L/g, V$_g$ is the gas flow rate which is 1 $\pm$ 0.05~L/min, t$_s$ is the time when the efficiency decrease to 95\% and m$_{AC}$ is the mass of activated carbon which is 0.73~g. At -60~$^\circ$C, the radon adsorption capability is 167.1 $\pm$ 2.7~L/g. The uncertainty is derived from V$_g$ and t$_s$.  The uncertainty of t$_s$ is the half of the bin width shown in Fig.~\ref{fig.charcoal60}. Uncertainty from m$_{AC}$ is negligible.

\begin{figure}[htb]
\centering
\includegraphics[width=7cm]{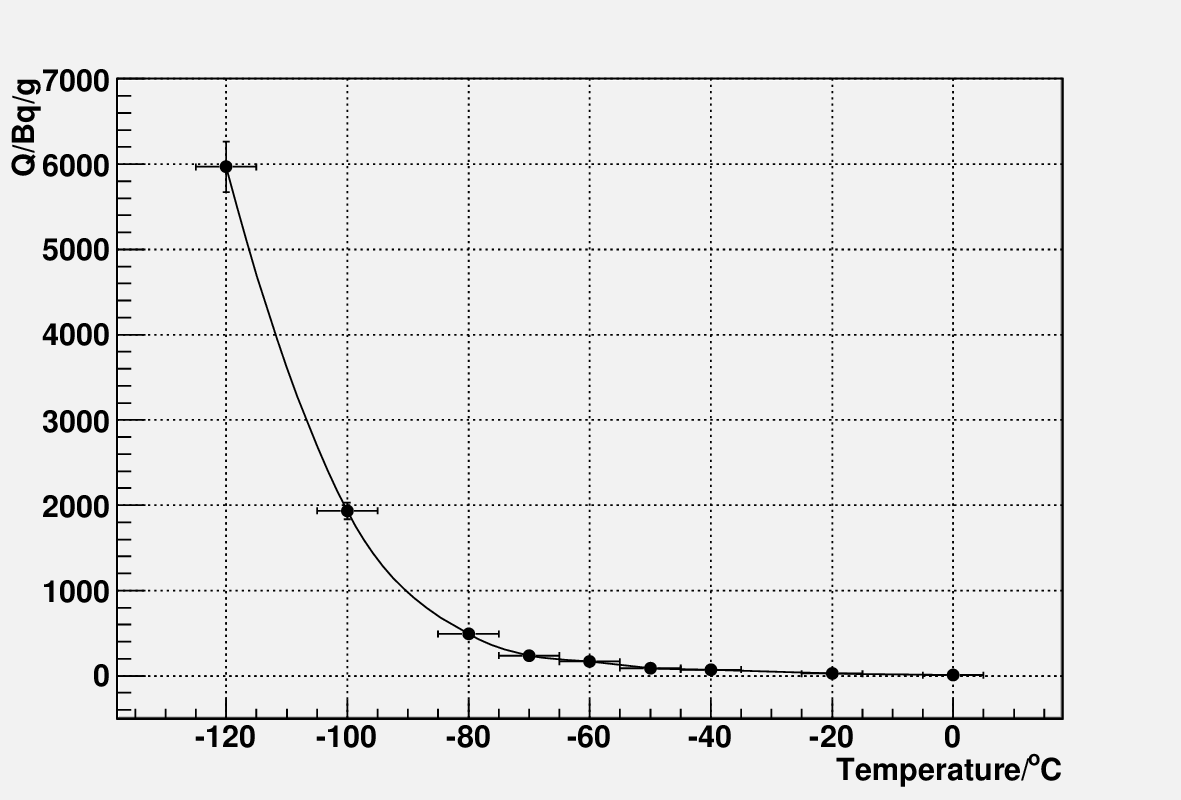}
\caption{Temperature dependence of adsorption capability. The uncertainty in the X-axis is the stability of the temperature, which is $\pm$5 ~$^\circ$C, the uncertainty in the Y-axis is derived from the stability of the gas flow rate and the time accuracy, the time accuracy is half of the bin width of Fig.~\ref{fig.charcoal60}.}
\label{fig.T_charcoal}
\end{figure}

The radon adsorption capability at different temperature has been studied with the same setup. Fig.~\ref{fig.T_charcoal} shows the temperature dependence of adsorption capability for Saratech activated carbon. The adsorption capability increases exponentially with decreasing temperature. At -120~$^\circ$C, 1~g activated carbon can be used to adsorb 5967 $\pm$ 298~L carrier gas (with radon) with an efficiency larger than 95\%. The radon concentration in the gas is $\sim$33~Bq/m$^3$.

\section{Conclusions and prospect}
Radon and radon daughters are crucial background sources for neutrino experiments and dark matter searching experiments. For the application of activated carbon in low background experiments, the intrinsic background and the radon adsorption capability are two main factors to be considered. A new radon-emanation measurement system, including a 0.71~mBq/m$^3$ high-sensitivity radon detector, and a new adsorption-performance research-system have been developed. The sensitivity of the radon detector reported in this work is very close to the number provided by Super-K group~\cite{SuperK3} which is the best in the world and our results also indicate that radon detectors with hemispherical geometry allow to both improve the collection efficiency and lower the operating voltage. In this paper, Saratech activated carbon has been studied in detail. The $^{226}$Ra concentration of the activated carbon is 1.36 $\pm$ 0.94~mBq/kg and its radon adsorption capability at -120~$^\circ$C is $\sim$6000~L/g.  With a further decrease of temperature, its radon adsorption capability is expected to be further enhanced. Saratech activated carbon is a good choice for radon removal and radon enrichment in the low background experiments.

\section{Acknowledgements}
This work is supported by Beijing Natural Science Foundation (Grant No. 1202024),  Yalong River Joint Fund of the National Natural Science Foundation of China and Yalong River Hydropower Development Co., LTD (Grant No. U1865208), the Innovative Project of Institute of High Energy Physics (Grant No. Y954514), the National Natural Science Foundation of China (Grant No. 11875280, No. 11905241) and the Xiejialin Foundation of Institute of High Energy Physics (Grant No. E05468U2).

\end{document}